# Stochastic Adaptive Single-Site Time-Dependent Variational Principle

Yihe Xu[†], Zhaoxuan Xie[†], Xiaoyu Xie[‡], Ulrich Schollwöck[§], Haibo Ma[†]

[†] School of Chemistry and Chemical Engineering, Jiangsu Key Laboratory of Vehicle Emissions Control, Nanjing University, Nanjing 210023, China.

[‡] Department of Chemistry, University of Liverpool, Liverpool L69 3BX, United Kingdom.

[§] Arnold Sommerfeld Center of Theoretical Physics, Department of Physics, University of Munich, Theresienstrasse 37, 80333 Munich, Germany and Munich Center for Quantum Science and Technology (MCQST), 80799 Munich, Germany

*Supporting Information Placeholder*

**ABSTRACT:** In recent years, the time-dependent variational principle (TDVP) method based on the matrix product state (MPS) wave function formulation has shown its great power in performing large-scale quantum dynamics simulations for realistic chemical systems with strong electron-vibration interactions. In this work, we propose a new stochastic adaptive single-site TDVP (SA-1TDVP) scheme to evolve the bond-dimension adaptively, which can integrate the traditional advantages of both the high efficiency of single-site TDVP (1TDVP) variant and the high accuracy of the two-site TDVP (2TDVP) variant. Based on the assumption that the level statistics of entanglement Hamiltonians, which originate from the reduced density matrices of the MPS method, follows a Poisson or Wigner distribution, as generically predicted by random matrix theory, additional random singular values are generated to expand the bond-dimension automatically. Tests on simulating the vibrationally-resolved quantum dynamics and absorption spectra in the pyrazine molecule and perylene bisimide (PBI) J-aggregate trimer as well as a spin-1/2 Heisenberg chain show that it can be automatic and as accurate as 2TDVP but reduce the computational time remarkably.

Based on a linear-chain matrix product state (MPS) representation for the many-body wavefunction with the advantages of high compression and local structure, the time-dependent variational principle (TDVP) approach [1,2] as a time evolution method for MPS (for other approaches see [3-9]) has been shown to be a powerful tool for simulating the quantum dynamics and spectroscopy of large realistic chemical systems with electron-vibration (electron-phonon) interactions [10-19].

In MPS-TDVP simulations, there are several types of errors, namely the projection error of projecting $\hat{H}|\psi\rangle$ onto the tangent space of the given MPS $|\psi\rangle$, the truncation error in tensor singular value decomposition (SVD) and the Krylov and time step error for the evolution operation of $e^{-i\hat{H}t}$ on $|\psi\rangle$.[3] The most widely used single-site TDVP (1TDVP) is symplectic and accordingly has zero truncation error, but the bond-dimension $m$ of the time-evolving MPS via 1TDVP cannot increase to accommodate entanglement increasing with time. In many cases, numerous test calculations need to be performed for a given task to test the convergence with respect to the bond-dimension $m$ and obtain a suitable value of $m$ for 1TDVP simulation. However, it may be quite tedious and expensive, especially for realistic complicated systems. To overcome this disadvantage, one can resort to two-site TDVP (2TDVP). Compared to 1TDVP, 2TDVP can adaptively increase the bond-dimension $m$ and has a smaller projection error but a non-zero truncation error. Generally, 2TDVP will be more accurate and robust, but less efficient than 1TDVP. Therefore, the current TDVP calculations of strongly correlated large systems always face a dilemma: it is difficult to balance the accuracy and robustness of 2TDVP and the computational efficiency of 1TDVP, especially for simulations of vibronic problems in which the size of local Hilbert spaces can be very large.

To break the above bottleneck, Yang and White [20] recently suggested a method to increase the bond-dimension of 1TDVP dynamically via temporarily creating an MPS representation of the current time-evolved state by expanding the MPS to represent both the current state and a sequence of Krylov vectors generated from it. Utilizing the benefit of the expanded manifold coming from this representation, the subsequent 1TDVP time step becomes more accurate and reliable. In addition, Dunnett and Chin [21] also proposed an autonomously adaptive variant of 1TDVP to capture the growing entanglement 'on the fly'. To ensure that the local bond-dimension $m$ can evolve, they utilized *full* QR factorization instead of the normally used *thin* or *reduced* QR factorization in updating the MPS local tensors. To measure the convergence behavior with increasing bond-dimension $m$, the norms of the effective Hamiltonian applied to their respective MPS site tensors are evaluated.

In this letter, we propose a new stochastic adaptive 1TDVP method to evolve the bond-dimensions automatically, which is computationally efficient without any additional tensor operations. This method is based on the statistical distribution law of tensor SVD singular values $\lambda$ (i.e. the square root of eigenvalues of the left or right subsystem's reduced density matrix) in MPS decomposition. It is well-known that the $\lambda$ spectrum shows a clear exponential decay as $\lambda_n \sim \exp(-\text{const} \times \ln^2 n)$ for the $n$th singular value in generic gapped one-dimensional systems, implying a linear relationship between the logarithmic singular values ($s_n \equiv \log(\lambda_n)$) and $\ln^2 n$ [22]. However, quantitative deviations from this relationship has been observed for large $m$ [23, 24], preventing the use of this property to estimate the necessary $m$ value for a pre-determined sufficiently small truncation threshold $\varepsilon$. There is, however, more information available: one can derive a so-called entanglement Hamiltonian [25-30] from the reduced density operator $\rho$ of an

MPS as $\hat{H}_{en} = -\log\rho$. Its eigenvalues (energy levels) are, up to a constant factor of no further importance, the $s_n \equiv \log(\lambda_n)$; their level spacings (or first-order differentials) are then given by $\Delta s_n = s_{n+1} - s_n$ and their second-order differentials by $\Delta_2 s_n = \Delta s_{n+1} - \Delta s_n$. We assume here that, in general, entanglement Hamiltonians exhibit the same level spacing statistics as many-body Hamiltonians do. There, one finds that the level spacings are distributed according to a Poisson distribution (as arises for randomly distributed eigenvalues) in the case of integrable systems and according to a Wigner distribution in the case of non-integrable systems, as a result of random-matrix theory [31-33]. This implies certain distributions (exponential and quasi-Gaussian) for the second-order differentials $\Delta_2 s_n = \Delta s_{n+1} - \Delta s_n$ (see supporting information), whose parameters are fitted from the spectrum of the reduced density operators. Using these distributions, we can estimate the small singular values in TDVP. The tests on simulating the vibrationally-resolved quantum dynamics and absorption spectra in the pyrazine molecule and perylene bisimide (PBI) J-aggregate trimer as well as spin-1/2 Heisenberg chain show that this can be automatized and be as accurate as 2TDVP, but save a lot of computational time.

To analyze the singular value distribution behavior in real-time TDVP simulations, we take a 4-mode exciton-vibration model for the pyrazine molecule [34] as an example,

$$\hat{H} = \sum_{i,j} \varepsilon_{ij} \hat{a}_i^\dagger \hat{a}_j + \sum_K \omega_K \hat{b}_K^\dagger \hat{b}_K + \sum_{i,j,K} g_{ij}^K \hat{a}_i^\dagger \hat{a}_j (\hat{b}_K^\dagger + \hat{b}_K). \quad (1)$$

Here, the three terms describe the electronic and vibrational parts as well as their interactions respectively. $\hat{a}_i^\dagger$ and $\hat{a}_i$ represent the creation and annihilation operators of electronic state $i$, while $\varepsilon_{ij}$ denotes the onsite energy of state $i$ ($i = j$) or the electronic coupling between state $i$ and state $j$ ($i \neq j$). $\hat{b}_K^\dagger$ and $\hat{b}_K$ are the raising and lowering operators for vibration mode $K$, while $\omega_K$ is the vibrational frequency of mode $K$. $g_{ij}^K$ is the coupling between vibrational mode $K$ and electronic Hamiltonian term. Due to the small number of the degrees of freedom in this model, we can perform an exact 2TDVP simulation (total time 4000 a.u., the time step is 20 a.u.) without any SVD truncation errors to avoid discarding small singular values. The computational details can be found in the Supporting Information. Then we extract the singular values $\{\lambda_n\}$ on each bond at different time steps, and show the logarithms of singular values $s_n$ versus $\log^2 n$ for two example SVDs with large bond-dimensions within short/long time ranges across multiple entanglement regimes in Figure 1a and Figure 1b.

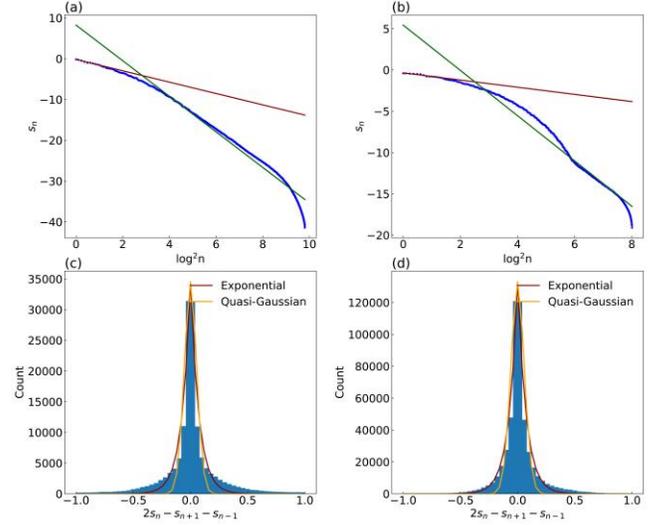

Figure 1. Singular value distribution versus $\log^2 n$ at (a) bond 4-5 at time point 300 a.u. and (b) bond 6-7 at time point 2000 a.u. The red and green lines are fittings by $y = -\text{const} \times \log^{1+1/\alpha} n$ with $\alpha = 1$ for the largest 10 singular values and all singular values respectively. Distributions of all the second-order differentials (c) at bond 4-5 during time range of 0-300 a.u. and (d) at bond 6-7 during time range of 0-2000 a.u. with exponential and quasi-Gaussian fittings.

There are obvious linear correlations for large $s_n$, i.e. small $n$, but the linear relationship breaks down for smaller $s_n$ (i.e. large $n$). This agrees with previous findings in grand canonical ensembles and Ising models [23, 24] and can be ascribed to the increasing degree of energetic degeneracy of higher excited states. However, interestingly, we find that $s_n$'s second-order differentials ($\Delta_2 s_n$) fulfill an exponential (Equation 2) or quasi-Gaussian (Equation 3) distribution, as shown in Figure 1c and Figure 1d.

$$P(\Delta_2 s) = \frac{\beta}{2} \exp(-\beta|\Delta_2 s|) \quad (2)$$

$$P(\Delta_2 s) = \beta^2 e^{-\frac{\beta}{4}(\Delta_2 s)^2} \times \int_{\left|\frac{\Delta_2 s}{2}\right|}^{+\infty} dy \left[ y^2 - \left(\frac{\Delta_2 s}{2}\right)^2 \right] e^{-\beta y^2} \quad (3)$$

This is related to random features in the vanishingly small singular values and the analogy between the occurrence of $s_n$'s first-order difference and inhomogeneous Poisson process or Wigner distribution (See supporting information for detailed discussion). It should be also noted that the exponential/quasi-Gaussian distribution behavior of the second-order differentials of logarithms of singular values may be general for other model Hamiltonians, such as the spin-1/2 Heisenberg model and Fermi-Hubbard model, as we show in Supplementary Figures S3 and S5.

Now we utilize the above exponential/quasi-Gaussian distribution behavior to implement the bond-dimension evolution in our stochastic adaptive 1TDVP (SA-1TDVP) method. Firstly, we recall the MPS wavefunction:

$$|\psi\rangle = \sum_{\sigma_1 \sigma_2 \dots \sigma_N} M^{\sigma_1} M^{\sigma_2} \dots M^{\sigma_N} |\sigma_1 \sigma_2 \dots \sigma_N\rangle, \quad (4)$$

where $\sigma_i$ represents the local $d$-dimensional basis on site $i$ and the $N$ is the total number of sites. $M(M^{\sigma_i}_{\alpha_{i-1},\alpha_i})$ are rank-3 tensors, where $\alpha_i$ represents the MPS bond-dimension index linking site $i$ and site $i$+1. To optimize the MPS wavefunction with a finite value of $m$ for the maximal bond-dimension, one needs to update $M^{\sigma_i}$ tensors successively. In the $i$th step of a 1TDVP left-sweep, only the tensor $M$ on single site $i$ is updated during the time evolution. In standard 1TDVP, a *reduced* SVD is then performed for the $(dm \times m)$-dimensional $M^{[i]}$ ($M_{\sigma_i \alpha_{i-1}, \alpha_i}$):

$$M^{[i]} = USV^\dagger. \qquad (5)$$

Here the $(dm \times m)$-dimensional matrix $U$ has orthogonal columns, and $S$ contains the $m$ singular values and square $(m \times m)$-dimensional $V^\dagger$ has orthogonal rows. To further run the time evolution at site $i$+1 and optimize $M^{\sigma_{i+1}}$ in the right-direction sweeping process, $M^{[i]}$ is then replaced by $U$, and $SV^\dagger$ is contracted into $M^{[i+1]}$. Because the number of non-zero singular values in $S$ is always no more than $m$, the bond-dimension of $M^{\sigma_i}$ will not exceed $m$. In order to increase the bond-dimension to describe the growing entanglement adequately, we extend the *reduced* SVD in Equation 5 to:

$$M^{[i]} = USV^\dagger \to (\,U \quad U_{\text{ex}}\,) \begin{pmatrix} S & 0 \\ O & 0 \end{pmatrix} \begin{pmatrix} V^\dagger \\ O \end{pmatrix}, \qquad (6)$$

where the block $U_{\text{ex}}$ is constituted from $\Delta m$ new column vectors. Gram-Schmidt orthogonalization is performed for the new randomly generated vectors in $U_{\text{ex}}$ with respect to the matrix $U$ from *reduced* SVD while the whole matrix ($U \quad U_{\text{ex}}$) is still column orthogonal but with shape $(dm \times (m + \Delta m))$. In the language of MPS, the columns in $U_{\text{ex}}$ represent the newly added bases in (left) subspace. The subspace expansion is realized by random generation and the subsequent Gram-Schmidt orthogonalization. $O$ is a block filled with 0. Note that if the sweep direction is reversed, one need to extend the matrix $V^\dagger$ instead.

To determine the column size $\Delta m$ of $U_{\text{ex}}$, we adopt a stochastic growth algorithm. The detailed SA-1TDVP algorithm for updating the bond-dimension of site $i$ at an evolved time step then works as:

1. Try 2TDVP for the initial $T$ (~20-50) time steps and save all the 2nd order difference values existing on each bond. Use the least square method (using mean-square error as the cost function) to fit their distributions by exponential/quasi-Gaussian lines (Equations 2 and 3, respectively) to get the parameter value for $\beta$. Here we assume that $\beta$ does not have a significant change during time evolution after the few initial steps, which is supported by Supplementary Figure S7. Then the fitting will be performed only once and the time cost of fitting is negligible compared with TDVP calculation.

2. Perform a *reduced* SVD on $(dm \times m)$-dimensional site tensor $M^{[i]}$ and obtain non-zero singular values $\{\lambda_n\}_{n \le m}$ and matrices $U, V^\dagger$. Set $N = m$.

3. Generate a new random number $\Delta_2 s_N$ satisfying the above exponential/quasi-Gaussian distributions with the parameter $\beta$ obtained from fitting and then calculate $\lambda_{N+1}$ from $\lambda_{N-1}, \lambda_N$ and $\Delta_2 s_N$.

4. If the new estimated singular value $\lambda_{N+1} > \lambda_N$ or $\lambda_{N+1} < \varepsilon$, exit the stochastic iteration process and go to step 5; otherwise add $\lambda_{N+1}$ to $\{\lambda_n\}_{n \le N}$ and set $N \leftarrow N + 1$; then go back to step 3.

5. Set $m_{\text{new}} = N$ and build $m_{\text{new}} - m$ new columns of $U$ (for left sweep) or $N$ rows of $V^\dagger$ (for right sweep) to update $M^{[i]}$.

Our approach for growing $m$ adaptively avoids the additional and computationally costly operation of applying the Hamiltonian operator to the MPS, whose complexity is $O(m^3 w d + m^2 w^2 d)$[3], where $w$ represents the bond-dimension of the matrix product operator (MPO). Moreover, as shown in Equations 5 and 6, the extension of the basis will not change the MPS itself, which is crucial for time evolution and isn't fulfilled by those perturbative methods suggested for static single-site DMRG calculations [35, 36]. We now first test the SA-1TDVP method by using the 4-mode pyrazine model. We compare the performances of convergence and accuracy of conventional TDVP and our method in Figure 2. The autocorrelation function $C(t)$ is defined as:

$$C(t) = \langle \psi(0) | \psi(t) \rangle. \qquad (7)$$

Here $|\psi(t)\rangle$ is the MPS at time $t$. Because the number of degrees of freedom in this model is small, we can perform an exact 2TDVP simulation with no truncation error and we take this as a reference. It is shown that one can obtain very accurate results using SA-1TDVP with an SVD truncation cutoff $\varepsilon$ of $10^{-8}$. The two different distribution formulas (exponential and quasi-Gaussian) give very similar performances. Figure 2b demonstrates that the error of time correlation functions for SA-1TDVP is of similar order of magnitude as the cutoff in the entire time evolution of this model. One may notice the slight increase of SA-1TDVP's errors at long-time limit. This inefficiency may be caused by the fact the added random vectors in SA-1TDVP are non-optimized and will be more likely orthogonal to the physically relevant renormalized states from 2TDVP when $m$ is large. Figure 2c further shows that with the same cutoff value SA-1TDVP costs 90% less time than 2TDVP. At the same time, as a stochastic algorithm, SA-1TDVP is found to generate nearly the same bond-dimension as 2TDVP as shown in Figure 2d. In this particular case the speedup of the two distributions is essentially the same, but this is a peculiarity. The efficiency gain for SA-1TDVP over 2TDVP was also found for the tests of Heisenberg model (see Supplementary Figure S4), where the quasi-Gaussian distribution shows a slightly better performance than the exponential one, saving 84% and 75% computational time when compared with 2TDVP respectively.

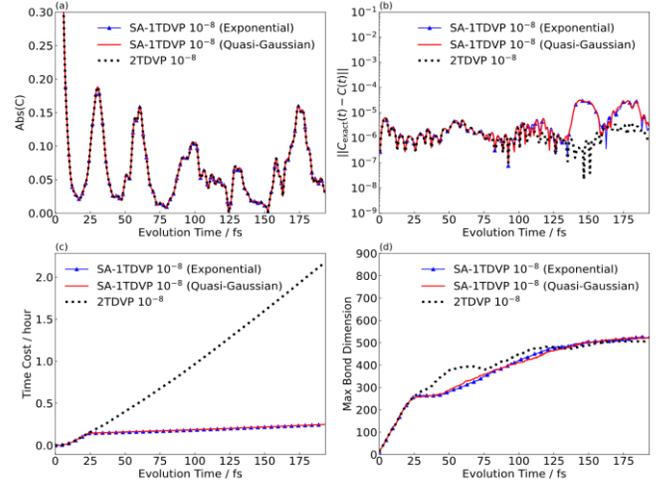

Figure 2. Results for 4-mode pyrazine model from SA-1TDVP ($T$=50) and conventional 2TDVP methods. (a) The absolute correlation function from conventional 2TDVP and SA-1TDVP. (b) The error of correlation functions, measured by the absolute value of differences between them with those by exact 2TDVP. (c) The time cost of the SA-1TDVP and 2TDVP. (d) The increase of max bond-dimension of SA-1TDVP and 2TDVP.

We now switch to the 24-mode pyrazine model [37] to show the performance of SA-1TDVP in larger systems. MPSs with different bond-dimension in conventional 1TDVP are obtained from a 1 a.u. preliminary 2TDVP calculation. For SA-1TDVP all dynamics start from an $m$=10 MPS. We find that one can obtain nearly converged results using SA-1TDVP with a decreasing SVD truncation cutoff $\varepsilon$, as shown in Figure 3a. From Figure 3b one can find that in the long-time range, the error of SA-1TDVP don't increase much because the adaption of bond-dimensions in the latter, as illustrated in Figure 3d. From Figure 3c and 3d we find that small change of cutoff may cause different behaviors. This is because finite cutoff and relatively large bond-dimension result in a dense sequence so that one can find many singular values in a small interval. Other computational details can be found in the Supporting Information.

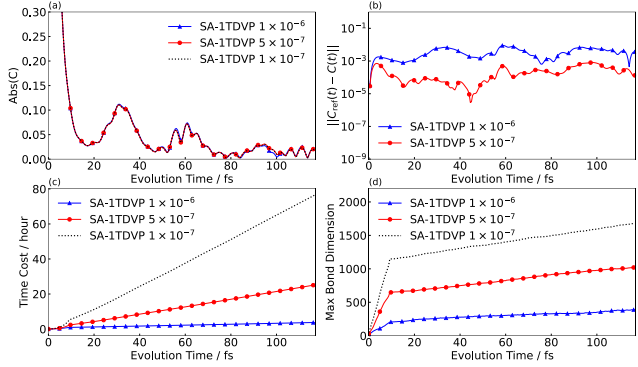

Figure 3. Results for 24-mode pyrazine model from SA-1TDVP with different cutoffs (a) The absolute autocorrelation function $C(t)$ from SA-1TDVP. (b) The error of correlation versus reference SA-1TDVP ($\varepsilon = 1 \times 10^{-7}$). (c) The time cost of the SA-1TDVP. (d) The increase of max bond-dimension of SA-1TDVP. All calculations use the exponential distribution here.

Finally, we test the calculation of the absorption spectra of 24-mode pyrazine and PBI trimer models [38, 39], and compare our SA-1TDVP results with other accurate 2TDVP and multi configuration time-dependent Hartree (MCTDH) [40] calculations in Figure 4. The SA-1TDVP results are in good agreement with accurate methods. Our method reproduces the details successfully, such as the small oscillations in Figure 4a and the relative magnitudes of peaks in Figure 4b. This implies that SA-1TDVP provides an efficient and automatic computational tool for the full quantum dynamics simulation of realistic chemical problems.

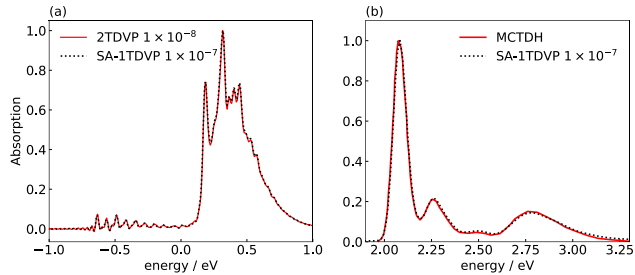

Figure 4. Simulated absorption spectra for (a) 24-mode pyrazine model and (b) PBI trimer model. Results of SA-1TDVP are compared with 2TDVP ($\varepsilon=10^{-8}$) [15] and MCTDH [38] reference results.

In conclusion, we observe an exponential/quasi-Gaussian distribution behavior for the second-order differentials of logarithmic singular values in DMRG and TDVP simulations for the first time, which can be ascribed to the Poisson or Wigner distribution of $\Delta s_n$. We propose a SA-1TDVP algorithm based on this distribution behavior to perform subspace expansion of MPS randomly and adaptively during the time evolution. We compare the accuracy and efficiency of our method in Heisenberg and exciton-vibration interaction models with conventional 2TDVP and 1TDVP as well as MCTDH. The tests on simulating the vibrationally-resolved quantum dynamics and absorption spectra in the pyrazine molecule and PBI trimer show that it can be automatic and as accurate as 2TDVP but increase the computational efficiency significantly. The automatic adaption makes that SA-1TDVP has the advantage of being able to avoid tedious tests of trying different bond-dimensions repeatedly in conventional 1TDVP. It is also worth to mention that, this strategy can be easily incorporated in any tDMRG code if there are SVD steps. For these reasons, SA-1TDVP provides an efficient, accurate and user-friendly tool for quantum dynamics simulations of large strongly correlated systems.

## ASSOCIATED CONTENT

### Supporting Information

The following files are available free of charge.
Discussion of the motivations for exponential and quasi-Gaussian distributions and computational details of 4-mode pyrazine, 24-mode pyrazine, PBI trimer and spin-1/2 Heisenberg and Fermi-Hubbard models as well as an illustrative application of SA-1TDVP for realistic novel chemical systems (exciton diffusion in a monolayer molecular crystal). (PDF)

## ACKNOWLEDGMENT

This work was supported by the National Natural Science Foundation of China (grant number 22073045) and the Fundamental Research Funds for the Central Universities and by the Deutsche Forschungsgemeinschaft (DFG, German Research Foundation) under Germany's Excellence Strategy-426 EXC-2111-390814868. We thank Xiangrong Wang, Martin Grundner and Sebastian Paeckel for stimulating discussions. We are also grateful to the High-Performance Computing Center of Nanjing University for carrying out the numerical calculations in this study on the blade cluster system.

# Supporting Information

Stochastic Adaptive Single-Site Time-Dependent Variational Principle

*Yihe Xu, Zhaoxuan Xie, Xiaoyu Xie, Ulrich Schollwöck, Haibo Ma*

## 1. Computational details

### 1.1 4-mode and 24-mode Pyrazine models

We first focus on the $S_1/S_2$ interconversion of pyrazine through a conical intersection between the $S_1$ and the $S_2$ states after UV photoexcitation to the $S_2$ state. This system is widely considered as an ideal model system for a rigorous test of the accuracy and capabilities of full quantum dynamics methods due to the strong vibronic couplings for both diagonal and off-diagonal terms. [S1] There are a widely used 24-mode model and a simplified 4-mode model for the pyrazine system. [S2, S3] Here, we present a brief introduction to them.

The equilibrium geometry of the ground-state pyrazine molecule conserves $D_{2h}$ symmetry, and the $S_1$ and $S_2$ states have symmetry $B_{3u}$ and $B_{2u}$ symmetry, respectively. Therefore, the symmetry of vibration mode (for linear electron-vibration coupling) or the product of symmetry of two modes (for bilinear coupling) should be $B_{1g}$ to make corresponding off-diagonal vibronic couplings nonzero. For the on-site coupling terms, this symmetry should be $A_g$. Consequently, the 4- and 24-model Hamiltonians can be generally expressed as,

$$\hat{H} = \begin{pmatrix} -\Delta & 0 \\ 0 & \Delta \end{pmatrix} + \sum_{I}^{4/24} \hbar\omega_I \left(\hat{b}_I^\dagger \hat{b}_I + \frac{1}{2}\right)$$
$$+ \sum_{I \in A_g} \hat{q}_I \begin{pmatrix} g_1^I & 0 \\ 0 & g_2^I \end{pmatrix} + \sum_{I \in B_{1g}} \hat{q}_I \begin{pmatrix} 0 & g_{12}^I \\ g_{12}^I & 0 \end{pmatrix}$$
$$+ \sum_{I \otimes J \in A_g} \hat{q}_I \hat{q}_J \begin{pmatrix} g_1^{IJ} & 0 \\ 0 & g_2^{IJ} \end{pmatrix} + \sum_{I \otimes J \in B_{1g}} \hat{q}_I \hat{q}_J \begin{pmatrix} 0 & g_{12}^{IJ} \\ g_{12}^{IJ} & 0 \end{pmatrix} \quad (S1).$$

Here, $\Delta$ is half the energy gap between $S_1$ and $S_2$ states. $I$ and $J$ are indices for vibration modes, and $g$ are electron-vibration couplings.

For the simplified 4-mode model ($A_g$ modes: $v_{6a}$, $v_1$, $v_{9a}$; $B_{1g}$ mode: $v_{10a}$), the last two terms in the equation containing second-order vibronic couplings are absent, and the parameters are extracted from reference [S2].

For the 24-mode model, both first and second-order interaction terms are involved, and all parameters can be found in the reference [S3].

To build the matrix product state (MPS) structure for time-dependent variational principle (TDVP) simulations in this work, vibration modes are represented by independent sites, and one site at the head of MPS represents all electronic states. Computational parameters such as time interval ($\Delta t$) and occupation number of each vibration mode ($N_{max}$) as well as the site ordering were tested for converging the final dynamics/spectroscopy results in our previous work [S4], and we adopted these optimal parameters of $\Delta t$ and $N_{max}$ for TDVP calculations in this work. For the site ordering, here we used a slightly different ordering (in this work we put all the electronic states into one site and put this site onto the end of the chain).

In our zero-temperature TDVP simulations, we employ the projected-purification (PP) algorithm [S5] to reduce the computational cost. We set the initial state as the product of $|S_2\rangle$ and all vibrational ground states. All TDVP simulations for the 4- and 24-mode pyrazine models are performed by using kln-x package [S6].

The simulated absorption spectra signal $I(\omega)$ can be written as Fourier transforms of time autocorrelation function $C(t)$ with a relaxation term:

$$I(\omega) = \int_0^t C(\tau) e^{i\omega\tau - \frac{\tau}{T}} \mathrm{d}\tau. \tag{S2}$$

Here $T$ is the relaxation time ($T$=200 fs in this section) and the time autocorrelation function $C(t)$ is defined as:
$$C(t) = \langle \psi(0)|\psi(t)\rangle. \tag{S3}$$

Here $|\psi(t)\rangle$ is the MPS at time $t$.

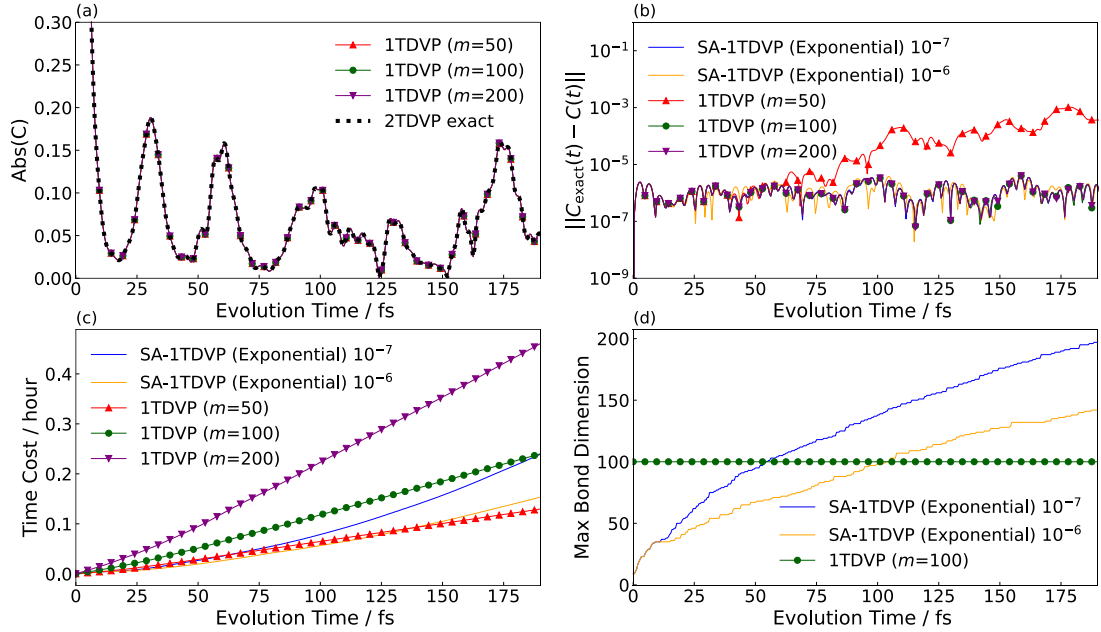

**Figure S1.** Results for 4-mode pyrazine model from 1TDVP and SA-1TDVP methods. (a) The absolute correlation function from 1TDVP and SA-1TDVP. (b) The error of correlation functions, measured by the absolute value of differences between them with those by exact 2TDVP. (c) The time cost of the 1TDVP and SA-1TDVP. (d) The increase of max bond-dimension of 1TDVP and SA-1TDVP. For a convenient comparison with 1TDVP, here all TDVP simulations were performed without using the PP method.

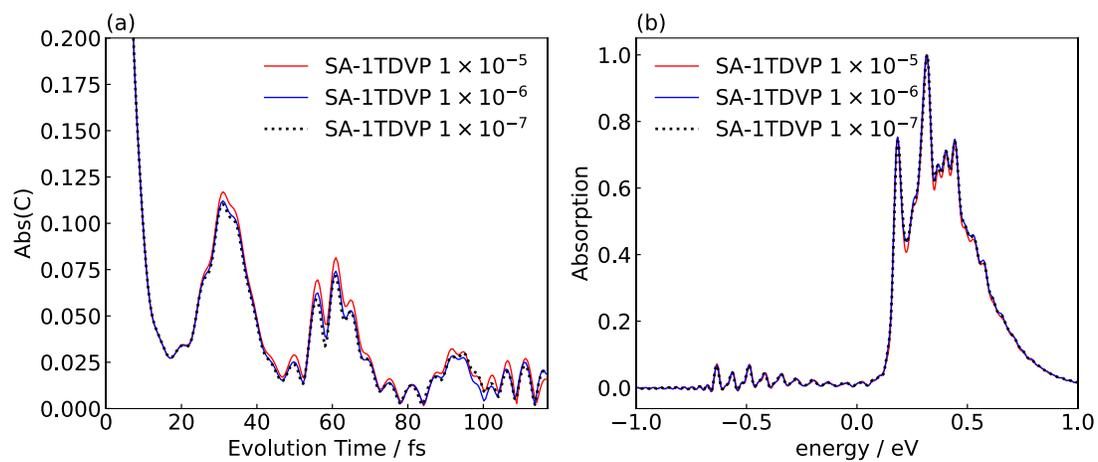

**Figure S2.** Convergence of (a) time autocorrelation function and (b) simulated absorption spectra for the 24-mode pyrazine model by SA-1TDVP versus different threshold parameters.

## 1.2 Spin-1/2 Heisenberg model

The Hamiltonian for the Heisenberg spin-$\frac{1}{2}$ chain can be written as:

$$\hat{H} = \sum_{i=1}^{N-1} \frac{J}{2}\left(\hat{S}_i^+ \hat{S}_{i+1}^- + \hat{S}_i^- \hat{S}_{i+1}^+\right) + J\hat{S}_i^z \hat{S}_{i+1}^z. \tag{S4}$$

Here $N$ is the number of sites of the chain and $J$ is the spin-coupling constant between neighboring sites. We choose $N = 24$ to perform 2TDVP for time length $10.0\, J^{-1}$ (step $0.1\, J^{-1}$) from a random initial state. And we capture the singular value distributions on the bond 12-13 during time evolution. Note that the singular values we only extracted belong to states whose spin z-components are 0. Like the Figure 1 in main text, we also show the singular value distribution in Figure S3 for the Heisenberg spin model.

From Figure S3 we can find that the small singular values deviate from the $\lambda_n \sim \exp(-\text{const} \times \ln^2 n)$ line like in the 4-mode pyrazine model. The distributions of second order differences are also fitted well by the exponential and quasi-Gaussian lines.

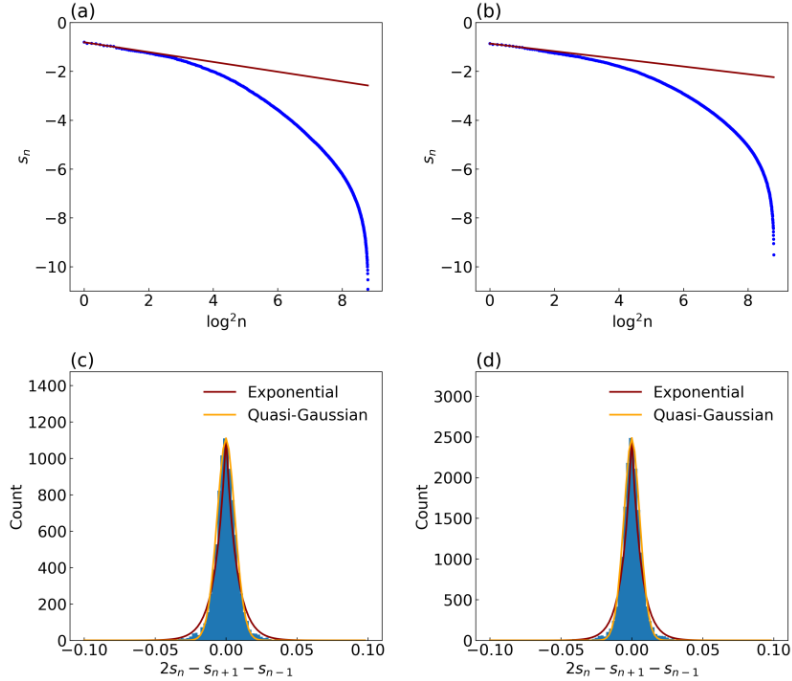

**Figure S3.** Singular value distribution versus $\log^2 n$ at (a) bond 12-13 time point $0.1\, J^{-1}$ and (b) bond 12-13 at time point $3.0\, J^{-1}$ in 24-site Heisenberg spin chain. The red line is fitted by $y = -\text{const} \times \log^{1+1/\alpha} n$ with $\alpha = 1$. (c) Distributions of second order differences at bond 12-13 at time point $0.1\, J^{-1}$. (d) Distributions of all second order differences at bond 12-13 during time range $0$-$3.0\, J^{-1}$.

Figure S4 shows the numerical performances of 2TDVP and SA-1TDVP. We use a 24-site Heisenberg chain with an initial Neel state to run dynamics simulation (total time $10.0\, J^{-1}$, step $0.1\, J^{-1}$, cutoff $10^{-6}$).

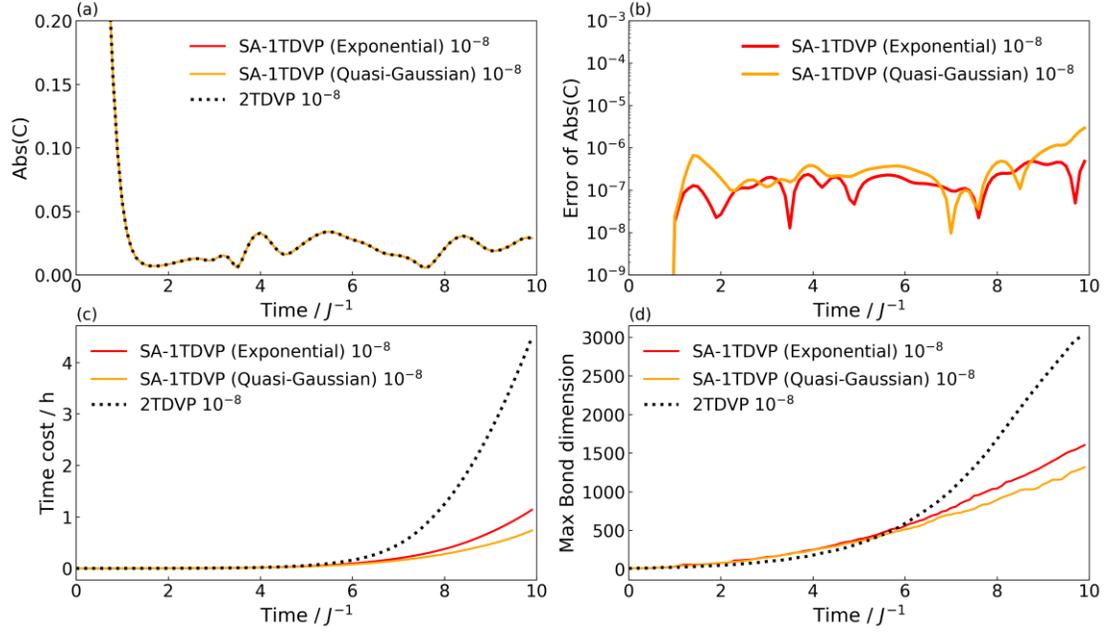

**Figure S4.** Results for 24-site Heisenberg spin model from SA-1TDVP and conventional 2TDVP methods. (a) The absolute correlation function from conventional TDVP and SA-1TDVP. (b) The error of correlation vs exact 2TDVP. (c) The time cost of the SA-1TDVP and 2TDVP. (d) The increase of max bond dimension of SA-1TDVP and 2TDVP.

## 1.3 Spin-1/2 Fermi-Hubbard model

We consider a Fermi-Hubbard model with spin-$\frac{1}{2}$ particles whose Hamiltonian can be written as:

$$\hat{H} = -\sum_{i=1}^{N-1} t\left(\hat{c}_{i,\uparrow}^+ \hat{c}_{i+1,\downarrow} + \hat{c}_{i+1,\uparrow} \hat{c}_{i,\downarrow}^+\right) + \sum_{i=1}^{N} U\left(\hat{c}_{i,\uparrow}^+ \hat{c}_{i,\uparrow} \hat{c}_{i,\downarrow}^+ \hat{c}_{i,\downarrow}\right). \tag{S5}$$

Here $N$ is the number of sites of the chain and $\hat{c}_{i,\uparrow}^+$, $\hat{c}_{i,\uparrow}$ are creation and annihilation operators for particles with spin up on site $i$, respectively. Both the length of model and particle numbers are 12. We then set the hopping integral $t$=1.0 and the repulsion term $U$=3.0. The total evolution time is 10.0 and step 0.1, starting from a random state. We extract the singular values in the central bond 6-7 and like the Heisenberg model mentioned above, we select those states with two conserved quantum numbers: the particle number 6 and total spin 0 in the blocks. Calculations in this section and data presented here was produced using the SyTen toolkit [S7], originally created by Claudius Hubig. The results are shown in Figure S5.

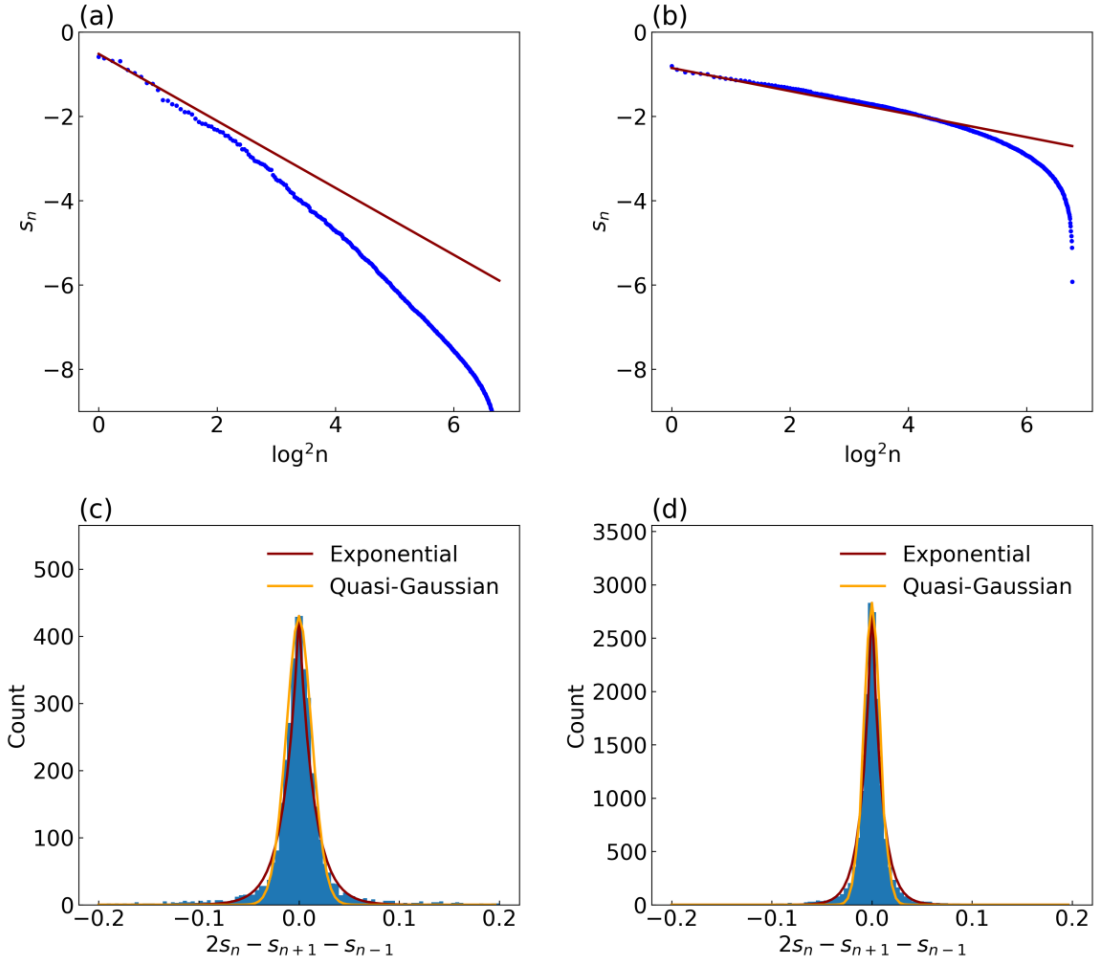

**Figure S5.** Singular value distribution versus $\log^2 n$ at the central bond (a) at time point 0.5 and (b) at time point 5.0. The red line is fitted by $y = -\text{const} \times \log^{1+1/\alpha} n$ with $\alpha = 1$. (c) Distributions of all second order differences during time range 0-0.5. (d) Distributions of all second order differences during time range 0-5.0.

## 1.4 PBI trimer model

The detailed Hamiltonian formula and its parameters in the J-aggregated PBI trimer model used for testing our method are extracted from ref. [S8]. We use max occupation number of 10 for all the vibrational modes [S9] and total evolution time 130 fs (step 0.13 fs). The trimer consists of three J-aggregated molecules. They have their own local $S_1$ and $S_2$ states. Here we consider two initial states: the superposition of three local $S_1$ and $S_2$ states, respectively. And the relaxation time of the two kinds of states are different. Therefore, the resulting spectra can be written as:

$$I(\omega) = \int_0^t d\tau \sum_{i=S1,S2} d_i^2 \left\langle i \left| e^{-i\hat{H}t} e^{-\frac{t}{T_i}} \right| i \right\rangle. \tag{S6}$$

Here the $\hat{H}$ is the Hamiltonian and $d_i$ is the transition dipole from $S_0$ to state $i$. We set the relaxation time $T_{S_1}$ = 30 fs and $T_{S_2}$ = 11 fs and dipole strength $d_{S_1}/d_{S_2}$=1.86.

## 2. Motivation of Exponential Distribution

As shown in Figure S6, we can regard the distribution of the first-order differential ($\Delta s \equiv s_{n+1} - s_n$) of logarithmic singular value ($s_n \equiv \log(\lambda_n)$ with $\lambda_n$ being the *n*-th singular value) as a Poisson process, if we assume $s_n$ are uniformly distributed in the whole range, which corresponds to homogeneous Poisson distribution.

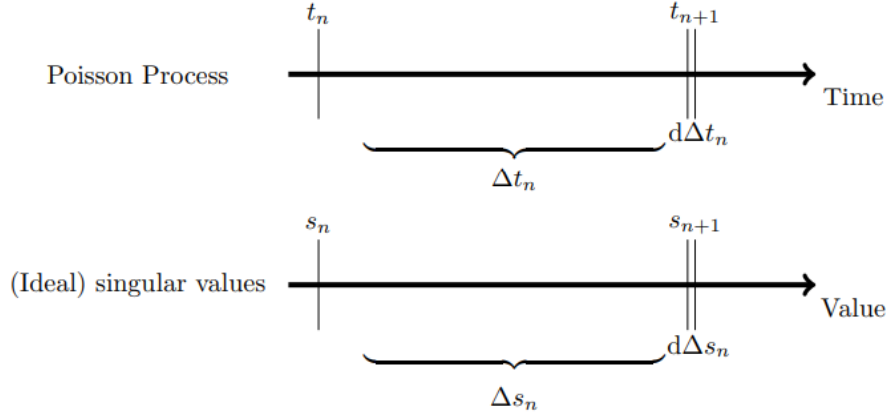

**Figure S6**. An analogy between singular values and Poisson process. One singular value found at certain point in "Value" axis behaves like "event" in Poisson process. "Ideal" means the singular values are uniformly distributed in the whole range, which corresponds to homogeneous Poisson distribution.

More specifically, in a Poisson process, the probability for the next event ($s_{n+1}$) to happen in the interval $[\Delta s_n, \Delta s_n + d(\Delta s_n)]$ after the last event ($s_n$) is

$$P(\Delta s_n)d(\Delta s_n) = \beta e^{-\beta \Delta s_n} d(\Delta s_n), \tag{S7}$$

with rate parameter $\beta$. Similarly, one has

$$P(\Delta s_{n+1})d(\Delta s_{n+1}) = \beta e^{-\beta \Delta s_{n+1}} d(\Delta s_{n+1}), \tag{S8}$$

and can define the second-order differential

$$\Delta_2 s = \Delta s_{n+1} - \Delta s_n. \tag{S9}$$

The probability of finding $\Delta_2 s$ to be in the interval $[\Delta_2 s, \Delta_2 s + d(\Delta_2 s)]$ is therefore the product of probabilities for $\Delta s_n$ and $\Delta s_{n+1}$:

$$P(\Delta_2 s)d(\Delta_2 s) = \int_0^{+\infty} d(\Delta s_n) \int_0^{+\infty \prime} d(\Delta s_{n+1}) \beta^2 e^{-\beta \Delta s_n} e^{-\beta \Delta s_{n+1}}. \tag{S10}$$

The prime denotes the constraint of $\Delta s_{n+1} \in [\Delta_2 s + \Delta s_n - d(\Delta_2 s), \Delta_2 s + \Delta s_n]$. To ensure $\Delta s_{n+1}(=\Delta s_n + \Delta_2 s)$ to be positive, the integral interval of $\Delta s_n$ is accordingly $[0, +\infty]$ for $\Delta_2 s \geq 0$ or $[|\Delta_2 s|, +\infty]$ for $\Delta_2 s < 0$. Therefore, we have

$$P(\Delta_2 s)d(\Delta_2 s) = \int_0^{+\infty} d(\Delta s_n)\beta^2 e^{-\beta \Delta s_n} e^{-\beta(\Delta_2 s + \Delta s_n)} d(\Delta_2 s)$$

$$= \beta^2 e^{-\beta \Delta_2 s} d(\Delta_2 s) \int_0^{+\infty} d(\Delta s_n) e^{-2\beta \Delta s_n}$$

$$= \frac{\beta}{2} e^{-\beta \Delta_2 s} d(\Delta_2 s), \tag{S11}$$

for $\Delta_2 s \geq 0$, and

$$P(\Delta_2 s)d(\Delta_2 s) = \int_{|\Delta_2 s|}^{+\infty} d(\Delta s_n)\beta^2 e^{-\beta \Delta s_n} e^{-\beta(\Delta_2 s + \Delta s_n)} d(\Delta_2 s)$$

$$= \beta^2 e^{-\beta \Delta_2 s} d(\Delta_2 s) \int_{|\Delta_2 s|}^{+\infty} d(\Delta s_n) e^{-2\beta \Delta s_n}$$

$$= \frac{\beta}{2} e^{\beta \Delta_2 s} d(\Delta_2 s), \tag{S12}$$

for $\Delta_2 s < 0$. We can finally combine Equations S8 and S9 to have an exponential distribution of arbitrary $\Delta_2 s$ as

$$P(\Delta_2 s)d(\Delta_2 s) = \frac{\beta}{2} e^{\beta |\Delta_2 s|} d(\Delta_2 s). \tag{S13}$$

It is easy to prove this probability function is normalized to 1, as

$$\int_{-\infty}^{+\infty} \frac{\beta}{2} e^{\beta |\Delta_2 s|} d(\Delta_2 s) = 2 \int_0^{+\infty} \beta e^{-\beta \Delta_2 s} d(\Delta_2 s) = 1. \tag{S14}$$

## 3. Motivation of Quasi-Gaussian Distribution

Besides the assumption of Poisson distribution for Δs in integrable systems, below we further consider the Wigner distribution for Δs in non-integrable systems. More specifically, the probability for the next event ($s_{n+1}$) to happen in the interval $[\Delta s_n, \Delta s_n + d(\Delta s_n)]$ after the last event ($s_n$) is

$$P(\Delta s_n)d(\Delta s_n) = \beta \Delta s_n e^{-\frac{\beta}{2}(\Delta s_n)^2} d(\Delta s_n), \tag{S15}$$

with rate parameter $\beta$.

Using the definition of the probability and integral interval in Equation S7, we have

$$P(\Delta_2 s)d(\Delta_2 s) = \beta^2 \int_0^{+\infty} d(\Delta s_n) \int_0^{+\infty} d(\Delta s_{n+1}) \Delta s_n \Delta s_{n+1} e^{-\frac{\beta}{2}(\Delta s_n)^2} e^{-\frac{\beta}{2}(\Delta s_{n+1})^2}$$

$$= \beta^2 d(\Delta_2 s) \int_0^{+\infty} d(\Delta s_n) \Delta s_n (\Delta s_n + \Delta_2 s) e^{-\frac{\beta}{2}(\Delta s_n)^2} e^{-\frac{\beta}{2}(\Delta s_n + \Delta_2 s)^2}$$

$$= \beta^2 d(\Delta_2 s) e^{-\frac{\beta}{2}(\Delta_2 s)^2} \int_0^{+\infty} d(\Delta s_n) [\Delta s_n \Delta_2 s + (\Delta s_n)^2] e^{-\beta[\Delta s_n \Delta_2 s + (\Delta s_n)^2]}$$

$$= \beta^2 d(\Delta_2 s) e^{-\frac{\beta}{4}(\Delta_2 s)^2} \int_0^{+\infty} d(\Delta s_n) \left[\left(\Delta s_n + \frac{\Delta_2 s}{2}\right)^2 - \left(\frac{\Delta_2 s}{2}\right)^2\right] e^{-\beta\left(\Delta s_n + \frac{\Delta_2 s}{2}\right)^2}$$

$$= \beta^2 d(\Delta_2 s) e^{-\frac{\beta}{4}(\Delta_2 s)^2} \int_{\frac{\Delta_2 s}{2}}^{+\infty} dy \left[y^2 - \left(\frac{\Delta_2 s}{2}\right)^2\right] e^{-\beta y^2} \tag{S16}$$

for $\Delta_2 s \geq 0$. But it is easy to generalize this quasi-Gaussian distribution for arbitrary $\Delta_2 s$:

$$P(\Delta_2 s)d(\Delta_2 s) = \beta^2 d(\Delta_2 s) e^{-\frac{\beta}{4}(\Delta_2 s)^2} \int_{\left|\frac{\Delta_2 s}{2}\right|}^{+\infty} dy \left[y^2 - \left(\frac{\Delta_2 s}{2}\right)^2\right] e^{-\beta y^2}. \tag{S17}$$

Unfortunately, the resulting formula is not closed, but involves essentially the Gaussian error function. In practice, one can derive the parameter and apply this distribution by tabulating Equation S17.

### 4. Fitting parameter versus time evolution

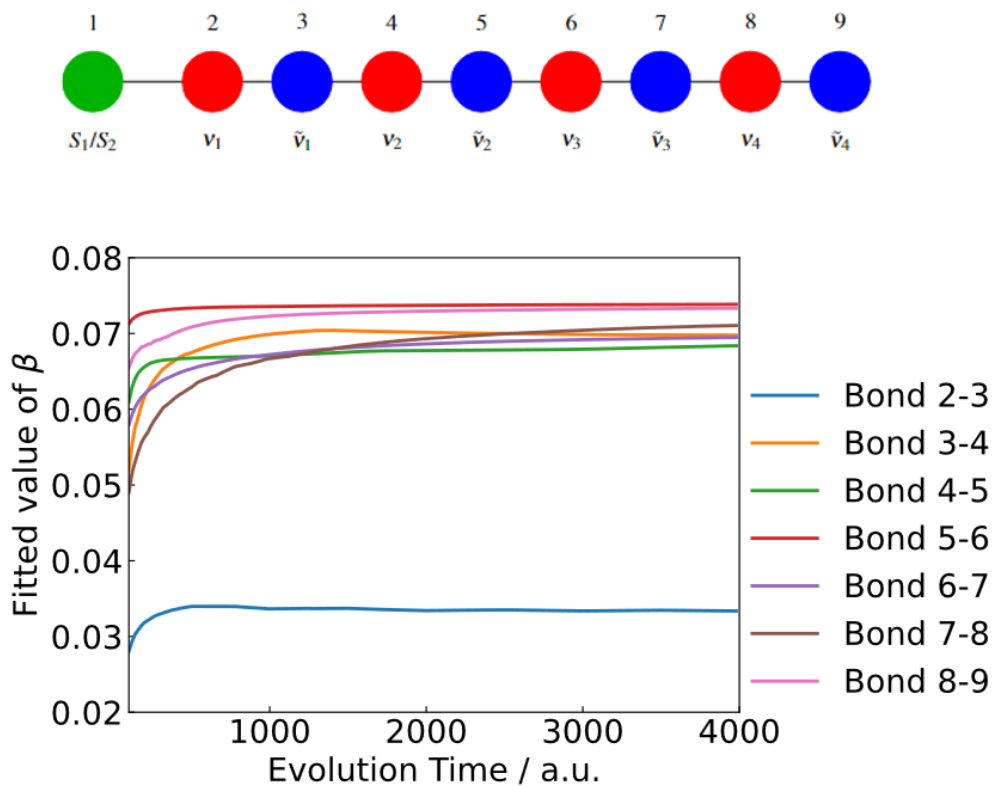

**Figure S7**. Fitted parameter $\beta$ in 4-mode pyrazine model for the exponential distribution from an exact 2TDVP simulation. Dimension of bond 1-2 is 2 all the time so we ignore it. $\tilde{v}_n$ represents the auxiliary mode corresponding to $v_n$.

## 5. Simulation of exciton diffusion in a monolayer molecular crystal

We build a 5 × 5 dimethyl-3,4,9,10-perylenetetracarboxilic diimide (Me-PTCDI) monolayer molecular aggregate, as shown in Figure S8a. To investigate the dynamics of the exciton diffusion in Me-PTCDI with SA-1TDVP, we still use the exciton-vibration coupling model:

$$\hat{H} = \varepsilon \sum_{n=1}^{25} |n\rangle\langle n| + J_{mn} \sum_{\langle mn \rangle}^{25} (|m\rangle\langle n| + |n\rangle\langle m|) + \omega_{\text{vib}} \sum_{n=1}^{25} b_n^\dagger b_n + \lambda \sum_{n=1}^{25} |n\rangle\langle n| (b_n^\dagger + b_n). \quad (S18)$$

Here $\varepsilon$ is the energy of the Frenkel exciton (FE) and $J_{mn}$ is coupling between FE states on nearest-neighbor monomers $m, n$. Only one effective vibrational mode with frequency $\omega_{\text{vib}}$ is considered on each monomer and its coupling with FE is $\lambda$. $|m\rangle$ means one FE exciton is on the monomer $m$. All these parameters are taken from ref.[S10]. We set the initial state as $|m = 13\rangle$, which is at the center of the whole aggregate. We then simulate the population dynamics of the system by using SA-1TDVP ($\varepsilon = 10^{-8}$), as shown in Figure S8. Here we use time interval ($\Delta t$) of 0.2 eV$^{-1}$ and set the maximum occupation number of each vibration mode ($N_{\text{max}}$) as 10.

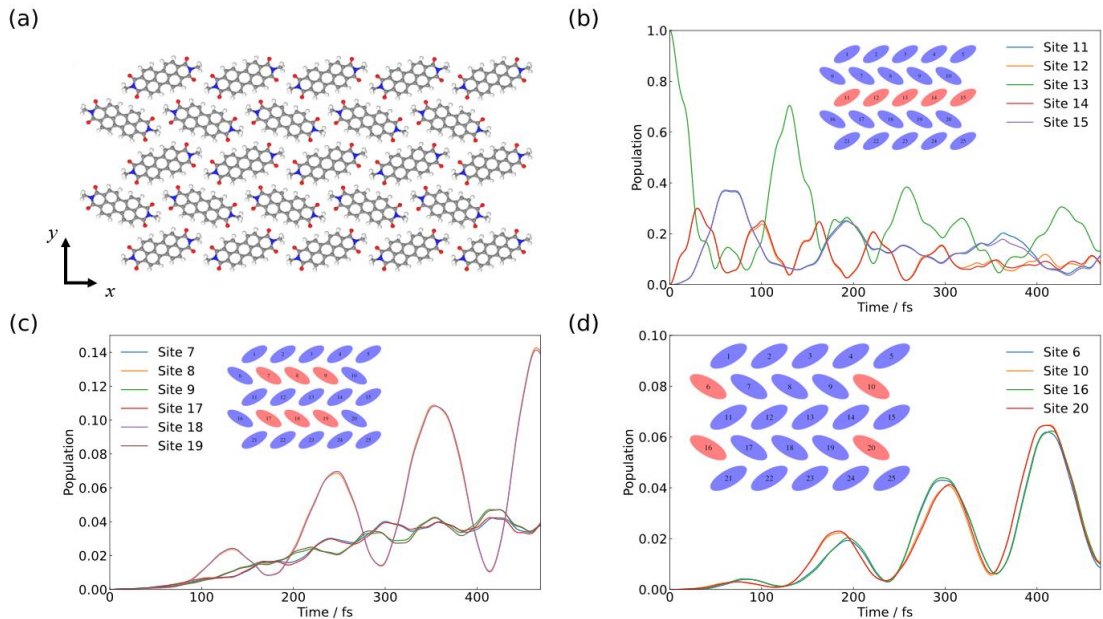

**Figure S8**. (a) Structure of ME-PCTDI for simulation. (b-d) population dynamics of selected sites.

We then calculate the correlation function and absorption spectra (See Equation S2 and Equation S3) as shown in Figure S9a and S9b. To characterize the delocalization of excitons, we extract the inverse participation ratio $P_R$ [S11] during time evolution:

$$P_R = \frac{1}{\sum_m \rho_m^2}. \tag{S19}$$

Here $\rho_m$ is the population of the FE state on the monomer $m$. The time evolutions of max bond dimension and $P_R$ are shown in Figure S9c and S9d.

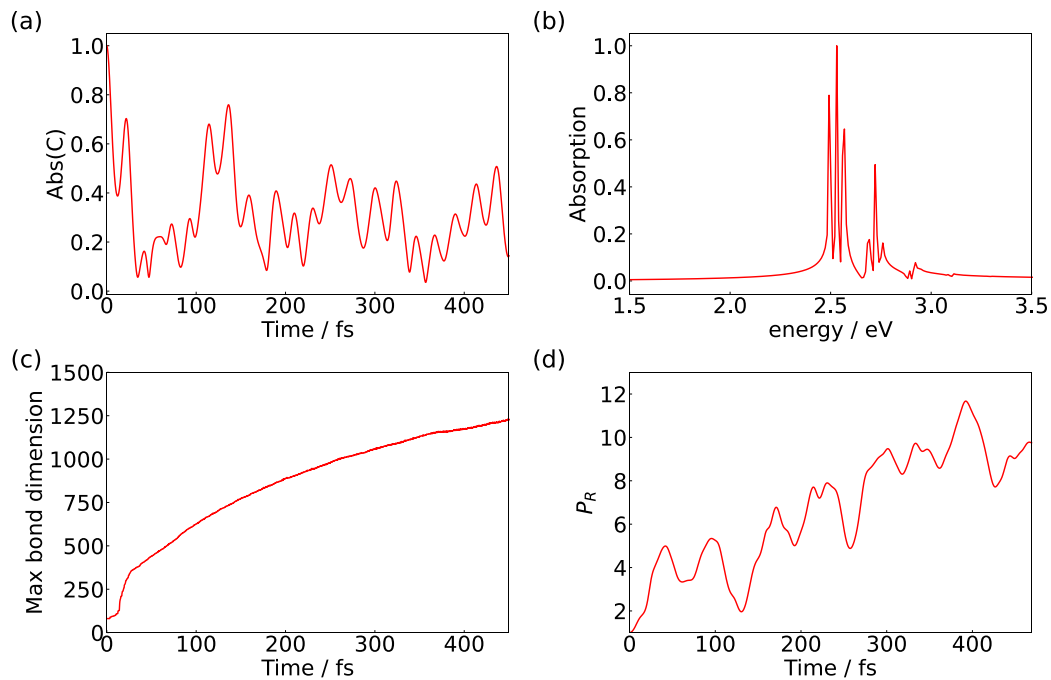

**Figure S9**. Results for the 25-molecule aggregate model from SA-1TDVP simulation. (a) The absolute time correlation function. (b) The absorption spectra. (c) Time evolution of max bond dimension. (d) Time evolution of inverse participation ratio.

We further analyze the phonon occupation. The populations of different occupation levels for all 25 vibrational modes are shown in Figure S10.

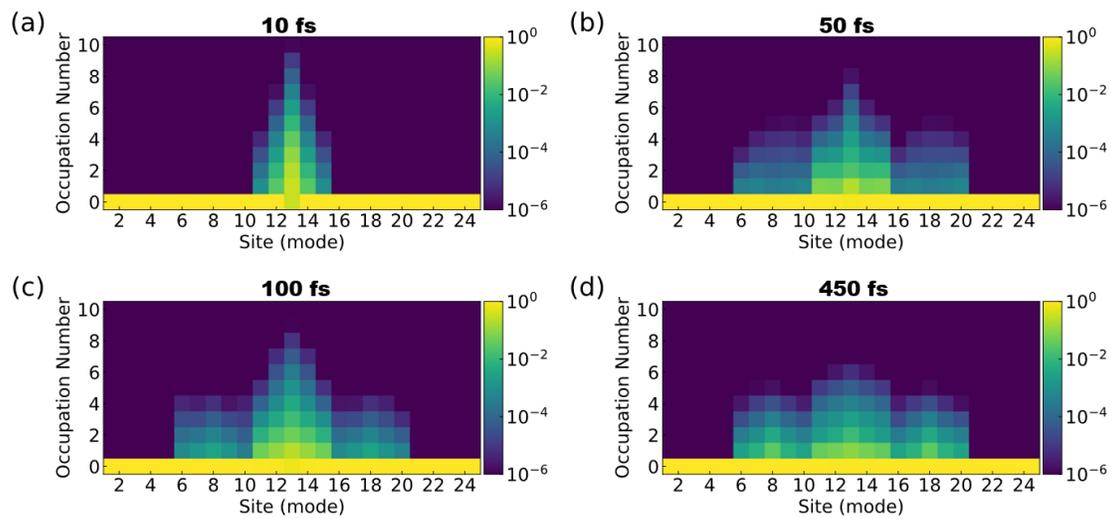

**Figure S10.** Calculated population of phonons on every occupation level on different vibrational modes at time point 10, 50, 100, 450 fs, respectively.